\documentclass[10pt,twocolumn,letterpaper]{article}

\usepackage{cvpr}              

\def\ds{$\mathcal{DS}\:$}

\def\tmr2{\multirow{2}} 
\def\tmr3{\multirow{3}} 
\def\tmc2{\multicolumn{2}}
\let\tb\textbf

\definecolor{cvprblue}{rgb}{0.21,0.49,0.74}
\usepackage[pagebackref,breaklinks,colorlinks,allcolors=cvprblue]{hyperref}
\usepackage{tcolorbox}
\usepackage{multirow}  
\newtheorem{definition}{Definition}

\def\paperID{7250} 
\def\confName{CVPR}
\def\confYear{2025}

\title{Mind the Gap: Detecting Black-box Adversarial Attacks in the Making through Query Update Analysis}

\author{Jeonghwan Park, Niall McLaughlin, Ihsen Alouani\\
Queen's University Belfast, United Kingdom\\
{\tt\small jeonghwan.park@uk.thalesgroup.com, \{n.mclaughlin, i.alouani\}@qub.ac.uk}
}

\begin{document}
\maketitle
\begin{abstract}
Adversarial attacks remain a significant threat that can jeopardize the integrity of Machine Learning (ML) models. In particular, query-based black-box attacks can generate malicious noise without having access to the victim model's architecture, making them practical in real-world contexts. 
The community has proposed several defenses against adversarial attacks, only to be broken by more advanced and adaptive attack strategies.  In this paper, we propose a framework that detects if an adversarial noise instance is being generated. Unlike existing stateful defenses that detect adversarial noise generation by monitoring the input space, our approach learns adversarial patterns in the input update similarity space. In fact, we propose to observe a new metric called Delta Similarity ($\mathcal{DS}$), which we show it captures more efficiently the adversarial behavior. We evaluate our approach against 8 state-of-the-art attacks, including adaptive attacks, where the adversary is aware of the defense and tries to evade detection. We find that our approach is significantly more robust than existing defenses both in terms of specificity and sensitivity.\footnote{Code is available at https://github.com/jpark04-qub/GWAD}

\end{abstract}   
\section{Introduction}
\label{sec:introduction}
Adversarial attacks have been investigated as a critical threat to the trustworthiness of ML systems. Two main threat models have been considered in the literature: white-box and black-box. The white-box setting assumes the adversary has access to the model parameters, architecture and gradient. This setting has been investigated thoroughly to provide a worst-case vulnerability analysis \cite{PGD_AT,FGSM,GOODFELLOW_BOOK}. 
%
\begin{figure}
    \centering
    \includegraphics[width=\linewidth]{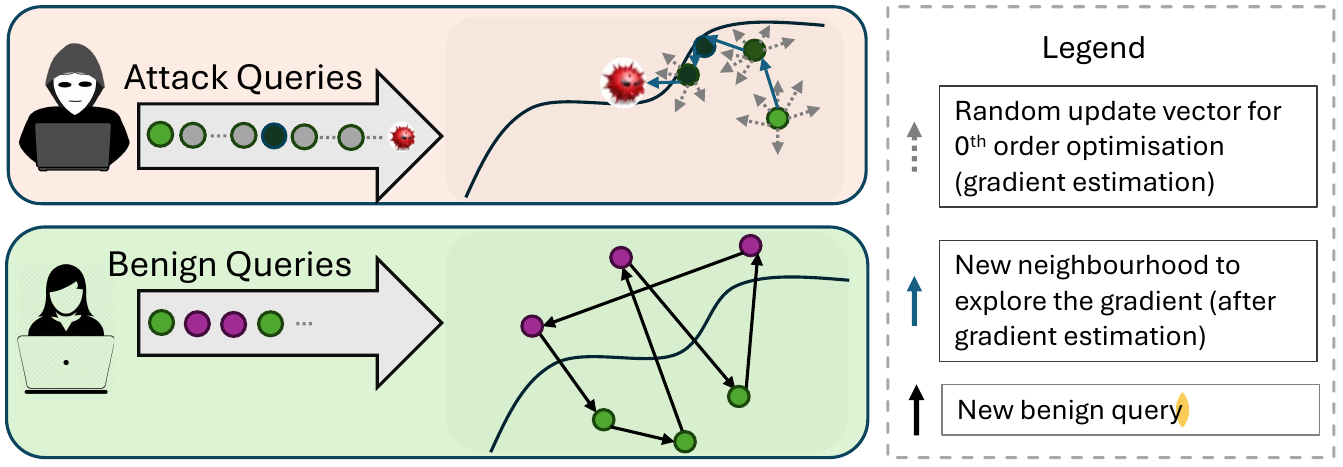}
   \caption{A high-level illustration of our intuition. The sequence of malicious queries to generate an adversarial example has a different pattern than  benign queries; Attack steps require \textbf{random vector updates} for gradient estimation.} 
   \label{fig:intuition}
\end{figure}
The second scenario, i.e., black-box, is more practical in real-world contexts and corresponds to an adversary with limited access to the victim model, restricted to inputs and outputs. This threat model is particularly relevant in Machine Learning-as-a-Service (MLaaS) settings, where models can be queried remotely by clients through APIs. A malicious actor generates adversarial perturbations by iteratively querying the model and analyzing the corresponding outputs. Despite the restricted access scenario, many query-based black-box attacks have been proposed \cite{HSJA, NES, SIGN_OPT, SIGN_FLIP}, showing a high efficiency in fooling ML models. 

Most existing defenses against adversarial attacks frame the problem in a "post-mortem" manner, i.e., assuming adversarial examples have already been generated \cite{PGD_AT,JEDI_PATCH,DAX_DEFENCE,EVAL_ADV_ROBUST}. However, a recent line of work specifically focused on \textit{query-based black-box settings} suggests to detect adversarial attacks in the making \cite{STATEFUL_DETECT,BLACKLIGHT,PIHA}. These approaches leverage the adversary's need to query the victim model, proposing to monitor input queries for anomalies in the input space that could indicate attempts to generate adversarial examples. For instance, Stateful Detection~\cite{STATEFUL_DETECT} identifies adversarial attack attempts by tracking the similarity (statefulness) between input queries, flagging anomalous queries that deviate significantly from typical behavior patterns. Similar methods, such as Blacklight~\cite{BLACKLIGHT} and PIHA~\cite{PIHA}, also rely on statefulness analysis. Blacklight focuses on detecting clusters of highly similar queries, which are often indicative of the iterative optimization techniques used by attackers. In contrast, PIHA uses statistical methods to analyze the input history, identifying anomalies in the distribution of queries.
All these approaches focus on the analysis of input samples, which can be evaded by adaptive attackers. In fact, many of these defenses have been shown to be vulnerable to an attack named Oracle-guided Adaptive Rejection Sampling (OARS) \cite{OARS}, which can bypass the defenses by carefully crafting queries that evade detection mechanisms. This vulnerability highlights the need for more robust defense strategies.

In this work, we tackle the detection of adversarial example generation in black-box settings from a new perspective, shifting the focus from input patterns to \textit{update patterns}. The primary intuition behind our approach is that query-based attack methods typically rely on variants of zeroth-order optimization to estimate gradients. 
The zeroth-order optimization estimates a gradient using multiple pairs of input queries and output observations. During this process, query-based attack methods iteratively update input examples in specific patterns and observe corresponding responses from the target model. Such structured updates are unlikely to be present in benign queries. \cref{fig:intuition} gives a generic illustration of our intuition. 
 Instead of relying on input space patterns, which sophisticated attackers can potentially evade, we introduce a novel indicator, termed \emph{Delta Similarity} ($\mathcal{DS}$), to differentially capture the relationship between updates in a sequence of queries over time. This metric captures more effectively the attack behavior, regardless of input examples or model architectures. By extracting features in the $\mathcal{DS}$ space, we train an ML model to identify attack patterns with high efficiency and low false positive rate.
In summary, our contributions are as follows:
\begin{itemize}
  \item We propose a novel metric, $\mathcal{DS}$, to analyze update patterns over time within a sequence of queries instead of focusing on the input examples. We show that $\mathcal{DS}$ is a reliable indicator of query-based adversarial attack behavior (Section \ref{Sec:delta_similarity}).
  \item Based on $\mathcal{DS}$, we propose a new framework called the Gradient's Watch Adversarial Detection (GWAD) to defend against query-based black-box attacks. GWAD includes an ML model trained on the \ds distribution and achieves high accuracy in adversarial attack detection with low false positive rate (Section \ref{Sec:methodoloty}).   
  \item We evaluate GWAD's generalization capabilities, showing that its performance is both dataset and model-agnostic, i.e., training GWAD on a particular model/dataset enables it to defend against attacks on different models and datasets (Section \ref{sec:generalisation}). 
  \item We evaluate our approach under two different adaptive attacks and show that GWAD is significantly more robust than state-of-the-art stateful defenses. Finally, we show that GWAD can be used in combination with a stateful defense to further enhance robustness (Section \ref{sec:enhancement}).
\end{itemize}

%
\begin{figure*}[t]
  \centering
   \includegraphics[width=1.0\linewidth]{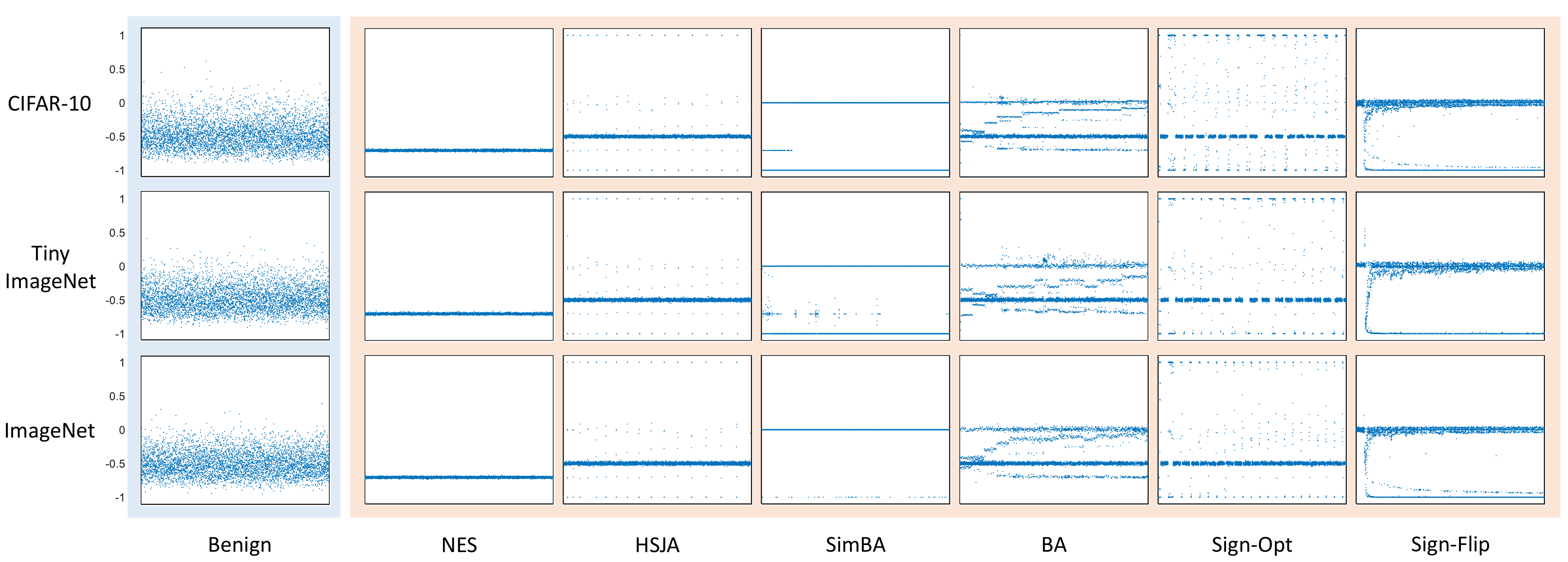}
   \caption{$\mathcal{DS}$ distribution of Benign and SOTA Query-based black-box adversarial attacks. We can see a clear difference between benign and adversarial attack distributions. Hence the distributions are amenable to classification.}
   \label{Fig:feature_distribution}
\end{figure*}
\section{Delta Similarity}
\label{Sec:delta_similarity}
Query-based black-box adversarial attacks commonly carry out $0^{th}$-order optimizations to estimate gradients, which consumes the majority of their attack queries. For example, HSJA \cite{HSJA} spends $96.06\%$ of its attack queries on $0^{th}$-order optimization, and remaining $3.93\%$ of the queries for linear optimization, e.g. binary search. The $0^{th}$-order optimization estimates a gradient using multiple pairs of input queries and output observations. During this process, query-based attack methods iteratively update input examples in specific patterns and observe corresponding responses from the target model.
Such updates are unlikely to exist in benign queries. 

\noindent \textbf{Theoretical insight:} Towards distinguishing the benign and malicious behavior, we are interested in analyzing the difference/similarity between queries in the input update space. Therefore, we define \ds as follows:
\begin{tcolorbox}[colback=blue!5!white,colframe=white]
\begin{definition} \textbf{(Delta Similarity)}
Let $\{x_{i-2}, x_{i-1}, x_{i}\}$ be three queries to a model $F$. Let $\delta$ be the variation between two queries, which we refer to as the update; $\delta$ is calculated by:
\begin{align}\label{Eq:delta_definition0}
\delta_{i} = x_{i} - x_{i-1}
\end{align}

We define \ds as the cosine similarity between updates $\delta_{i-1}$ and $\delta_{i}$; \ds is expressed by: 
\begin{align}
    \mathcal{DS} = \cos \angle\delta_{i-1}\delta_i
\end{align} \vspace{-6mm}
\label{def::ds}
\end{definition}
\end{tcolorbox}

In query-based black-box attack settings, each iteration involves estimating a gradient, which is used to optimize the current intermediate adversarial example $\tilde{x}_{t}$, pushing it toward maximizing the loss function of the victim model. Therefore, $\tilde{x}_{t}$ is not updated to $\tilde{x}_{t+1}$ until the gradient estimation step is completed. We denote the set of queries to the target model during a gradient estimation step as follows:
\begin{equation}
  \{ \tilde{x}_{t} + u^{j}_{t},\; \tilde{x}_{t} + u^{j+1}_{t},\; \tilde{x}_{t} + u^{j+2}_{t},\; ...,\; \tilde{x}_{t} + u^{n}_{t} \}
\label{Eq:delta_definition1}
\end{equation}
where each $u^{j}_{t} \in \mathbb{R}^{d}$ represents a $d$-dimensional random vector, sampled from a normal distribution $\mathcal{N}(0, \sigma^{2})$, and $n$ indicates the last query to the target model in the iteration.
The variations $\delta$ between examples are found by applying \cref{Eq:delta_definition0} to the examples as: 
\begin{align}
    &\delta^{j}_{t} \quad = (\tilde{x}_{t} + u^{j+1}_{t}) - (\tilde{x}_{t} + u^{j}_{t}) \quad = u^{j+1}_{t} - u^{j}_{t} 
\label{Eq:DS_type0}    
\end{align}
\noindent \textbf{Observation:} As shown in \cref{Eq:DS_type0}, the variation $\delta^{j}_{t}$ and $\delta^{j+1}_{t}$ are expressed as a linear operation between random vectors $u^{j}_{t}, u^{j+1}_{t},$ and $u^{j+2}_{t}$. 
Due to the concentration of measure phenomenon~\cite{CONCENTRATION_OF_MEASURE}, these random vectors are expected to be orthogonal in the hyper-dimensional space and to have equal length with a very high probability~\cite{BANDEIRA_LECTURE_NOTES}. \\
\noindent \textbf{Hypothesis:} Since this observation is satisfied by the $0^{th}$-order optimization process, \ds, as defined in \cref{def::ds}, represents specific patterns for query-based adversarial sequences, that are different from benign \ds. 
In the Supplementary Material (Section 1), we provide further geometric illustrations of this hypothesis. In the following, we present an empirical study to verify its plausibility. \\ 
\noindent\textbf{Empirical analysis:} We want to differentially observe the distribution of \ds for benign and malicious queries. We conduct an empirical analysis with various sequences of queries from benign examples as well as SOTA black-box attacks. The analysis is carried for three settings; each setting corresponds to an image classification dataset along with a target model. Specifically, we use ResNet-18, EfficientNet, and VGG-16 trained on CIFAR-10, Tiny-ImageNet and ImageNet, respectively. 

The \ds distribution of the benign queries is calculated by injecting $5K$ benign examples from the test dataset. In contrast, the \ds distribution of attack queries is acquired by performing query-based attacks over random samples from the dataset under the same $5K$ query budget. 
The \ds distributions shown in \Cref{Fig:feature_distribution} suggest that the query-based black-box adversarial attack methods leave unique traces in the \ds space during their attack processes. Note that while the \ds of benign examples are distributed with a high variance, the \ds distribution of the attacks have one or more primary components. For example, NES \cite{NES} exhibit a strong \ds component at $-0.7071$, and HSJA \cite{HSJA} shows most of \ds at $-0.5$ and $\pm1.0$. More detailed rationale behind these numbers is provided in the Supplementary Material (Section 2). The patterns of adversarial attacks' \ds distributions are highly distinguishable from the ones in a sequence of queries of benign examples. This supports our hypothesis on the attack behavior distinguishability in the \ds space. 
\section{Proposed defense framework: GWAD}
\label{Sec:methodoloty}
\begin{figure}[t]
  \centering
    \includegraphics[width=1.0\columnwidth]{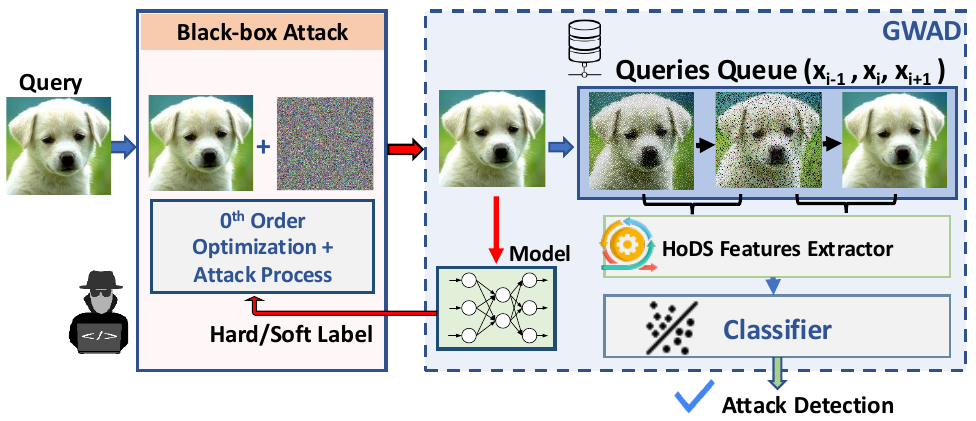}
   \caption{Block diagram of the procedure of GWAD query-based adversarial attack detection framework.}
   \label{Fig:GWAD_block_diagram}
\end{figure}
\begin{figure}[t]
    \centering
\includegraphics[width=1.0\columnwidth]{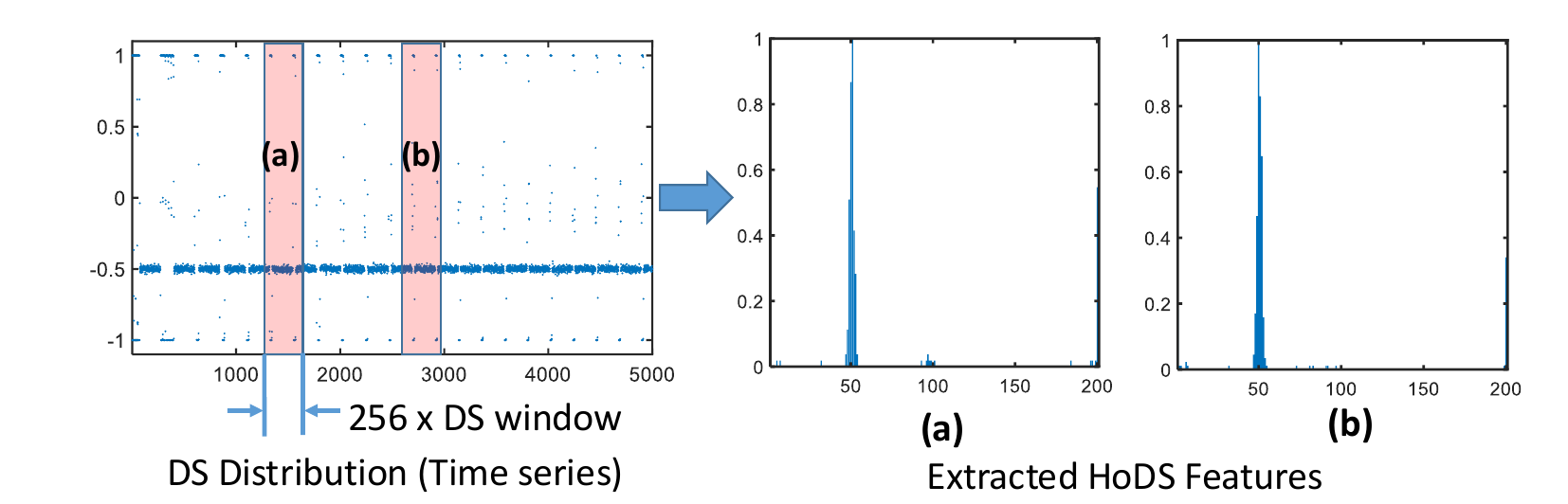}
   \caption{ \ds distribution of $5K$ attack queries from Sign-OPT \cite{SIGN_OPT} (Left);  HoDS corresponding to 2 windows--(a),(b)-- (Right).}
   \label{Fig:HoDS_feature_visualise}
\end{figure}
The main insight from the previous section 
is that the behavior of query-based black-box attacks can be characterised by the patterns in \ds distributions.
We propose a framework to identify ongoing adversarial generation attempts and detect adversarial examples in the making. Based on the properties of \ds, we propose a detection framework named Gradient Watch Adversarial Detection (GWAD). The proposed framework is illustrated in \Cref{Fig:GWAD_block_diagram}; it consists of continuously monitoring the \ds of input queries and analyzing the \ds~distribution using an ML model to detect if the protected system is under attack. 

\subsection{Features for Attack Detection/Classification}
\label{Sec:feature}
\ds distributions of attacks are not only highly distinguishable from benign queries, but each attack method also exhibits its own unique characteristics. Therefore, we capture a set of calculated \ds, and use the histogram representation to generate features for the attack classification problem as illustrated in \Cref{Fig:HoDS_feature_visualise}. In the remainder of this paper, we refer to these features by ``Histogram of \ds" (HoDS). 

For every triplet of queries we calculate a \ds. We generate a histogram of the last $256$ \ds, which is empirically chosen for the best prediction as detailed in the Supplementary Material (Section 6). \ds is bounded by $-1.0 \leq \mathcal{DS} \leq 1.0$. The HoDS has $200$ equally spaced bins over this range, with an extra bin for $\mathcal{DS} = 1.0$. The histogram's intensity varies over the sequence of queries, making local contrast normalization necessary; we apply min-max normalization for this purpose. Finally, a $1 \times 201$ HoDS feature vector is generated.

\subsection{Attack Classifier}
\label{Sec:deltanet_arch}
In the proposed framework, the extracted HoDS feature vector is input into a classifier to primarily detect attack queries and, ultimately, to identify the specific attack method. For this purpose, we train a fully-connected neural network comprising six fully-connected layers with ReLU activation functions. Classification probabilities are generated by a Log-Softmax output layer. 

The network architecture is designed to be as lightweight as possible following a brief hyperparameter optimization.We train our network for $100$ epochs with Stochastic Gradient Descent (SGD) optimizer and a batch size of $128$. 
\section{Experiments}
\label{sec:experiments}
\begin{table}[tp]  
  \centering
  \begin{tabular}{l|ccc}
    \toprule			
    {Dataset} & {Model}  & {Accuracy}  & {Usage}\\
    \midrule
    \multirow{2}{*}{CIFAR-10} & {ResNet-18}       & 93.07\% & {Train/Test} \\	
                              & {MobileNet-V2}    & 93.90\% & {Evaluation} \\	
    \midrule
    \multirow{2}{*}{ImageNet} & {VGG-16}          & 78.98\% & {Train/Test} \\
                            & {EfficientNet-b3}   & 83.58\% & {Evaluation} \\	
    \bottomrule                        
  \end{tabular}
  \caption{DNN Models' accuracy. Some DNNs are used to train GWAD while others are used for attack detection evaluation} 
  \label{Tab:DNN_acuracy}      
\end{table}
\begin{table}[t]
  \centering
  \footnotesize
  \begin{tabular}{c|ccc}
    \toprule
     {Categories} & {Attack Method} & {ASR} & {Update Type}  \\
     \midrule
     \multirow{3}{*}{Soft-Lable} & \textbf{NES}\cite{NES}            & 98.40\% & Normal\\
                                 & \textbf{SimBA}\cite{SIMBA}        & 97.60\% & One Hot Pixel\\
                                 & \textbf{BA}\cite{BA}              & 50.90\% & Normal\\
     \midrule
     \multirow{3}{*}{Hard-Label} & \textbf{HSJA}\cite{HSJA}          & 99.50\% & Normal, Uniform\\
                                 & \textbf{Sign-Opt}\cite{SIGN_OPT}  & 99.30\% & Normal\\
                                 & \textbf{Sign-Flip}\cite{SIGN_FLIP}& 95.70\% & Normal, Uniform\\ 
     \bottomrule                        
  \end{tabular}
  \caption{Query-based Black-box Adversarial Attack Methods. Their baseline ASR (untargeted) on MobileNet-V2 trained on CIFAR-10 dataset, and the type of update.} 
  \label{Tab:list_attack_methods}  
  \end{table}

%
To evaluate the performance of the proposed framework, we conduct experiments in three phases:
\textbf{(i)} the generalization capacity across different image classification tasks and their respective DNN models;
\textbf{(ii)} the attack detection performance, including the false positive rate across various practical use cases;
\textbf{(iii)} the robustness against potential adaptive attack scenarios.
\subsection{Experimental Setup}
\label{sec:experiment_setup}
We apply GWAD to scenarios in which DNN models for image classification tasks are targeted by query-based black-box attack methods. 

\noindent\textbf{Datasets and models: } To evaluate the effectiveness of GWAD, we use two standard image classification datasets: CIFAR-10 \cite{CIFAR-10} and ImageNet \cite{IMAGENET-1000}. Additionally, we consider three practical use cases with datasets chosen for their diverse data sample similarities to simulate varying challenge levels: specifically, the Hollywood Heads Dataset \cite{HOLLYWOOD_HEADS}, FLIR ADAS \cite{FLIR_ADAS_V2}, and BIRDSAI \cite{BIRDSAI}. The DNN models and their baseline accuracies are presented in  \Cref{Tab:DNN_acuracy}. 

\noindent \textbf{Attack configurations : } We evaluate the attack detection performance on six SOTA query-based black-box attack methods, as listed in \Cref{Tab:list_attack_methods}. Adversarial examples are generated using samples from the test set, with each attack method configured to operate within a $5K$ query budget. The ratio of perturbation added by the attack methods to achieve a successful attack is calculated as $\rho = \| \tilde{x} - x \|_{2} / \| x \|_{2}$, where $\tilde{x}$ is an adversarial example computed by the attack \cite{SQBA}. The attack is limited with a noise ratio budget $\rho = 0.1$. 
The hyper-parameters of the attacks methods used in this experiments are detailed in the Supplementary Material (Section 5). 

\begin{figure*}[t]
  \centering
   \includegraphics[width=\linewidth]{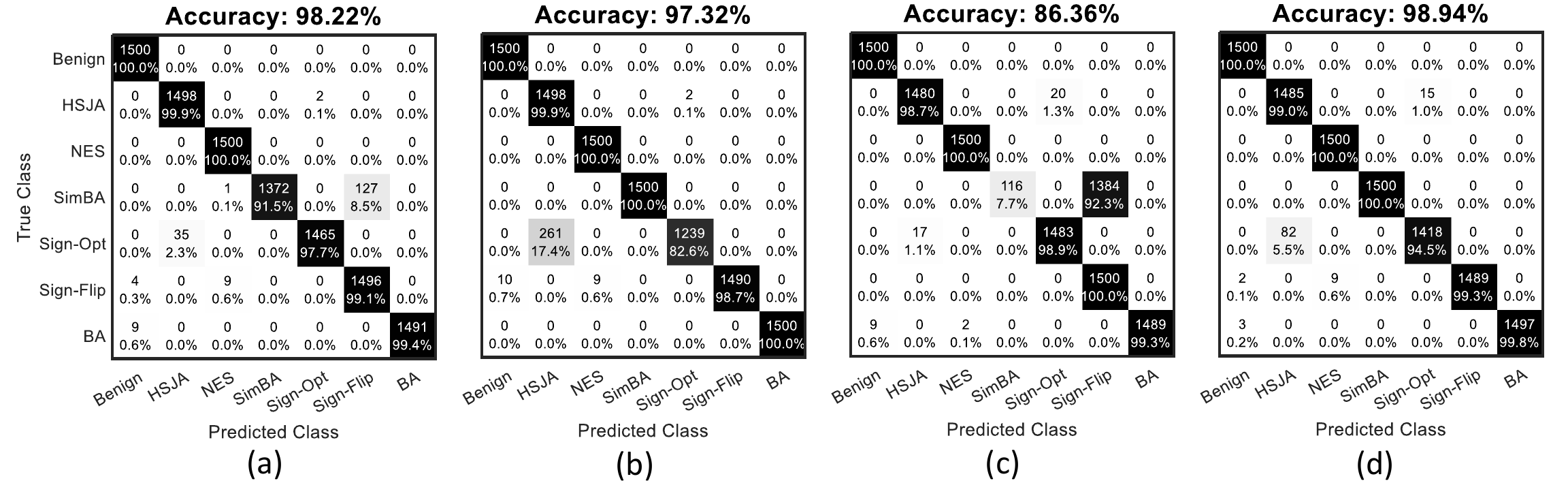}
   \caption{Confusion matrix of GWAD attack classification performance over validation HoDS feature sets: (a) and (b) show GWAD-CIFAR10 performance over CIFAR-10, and ImageNet, respectively; (c) and (d) show GWAD-ImageNet performance over CIFAR-10, and ImageNet, respectively.}
   \label{Fig:HoDS_generalisation}
\end{figure*}
\subsection{Generalization Property Evaluation}
\label{sec:generalisation}

In this section, we aim to evaluate GWAD's generalization performance across datasets—that is, if GWAD is trained on the \ds distribution of one dataset, can it generalize to detect attacks on other datasets? To test this, we extract HoDS features and train GWAD on two datasets: CIFAR-10 and ImageNet. For clarity, we refer to these trained classifiers as GWAD-CIFAR10 and GWAD-ImageNet. Note that the selection of these state-of-the-art attacks is intended solely to demonstrate GWAD’s ability to recognize attack patterns, as outlined in our preliminary analysis (\Cref{Sec:delta_similarity}).
\begin{table}[t!]
\centering
\begin{tabular}{lcccc}
\hline
\multirow{2}{*}{Attack} & \multicolumn{2}{c}{ImageNet} & \multicolumn{2}{c}{CIFAR-10} \\ \cline{2-5} 
                                  & {Recogn.} & {Detect.}   & {Recogn.} & {Detect.}   \\ \hline
HSJA                              & 90.71\%          & 100.00\%          & 98.12\%          & 99.99\%           \\
NES                               & 100.00\%         & 100.00\%          & 100.00\%         & 100.00\%          \\
SimBA                             & 100.00\%         & 100.00\%          & 5.60\%           & 100.00\%          \\
Sign-Opt                          & 62.53\%          & 99.92\%           & 98.34\%          & 100.00\%          \\
Sign-Flip                         & 93.68\%          & 99.42\%           & 99.26\%          & 100.00\%          \\
BA                                & 88.17\%          & 99.70\%           & 99.52\%          & 99.56\%           \\ \hline
\end{tabular}
\caption{GWAD sensitivity to the sequences of 100K attack queries against EfficientNet trained on ImageNet and MobileNet-v2 trained on CIFAR-10. "Recogn." corresponds to the attack recognition (multiclass, benign + 6 attacks), and the "Detect" corresponds to the binary classification (benign/attack).}
\label{Tab:gwad_sensitivity}
\end{table}
GWAD-CIFAR10 and GWAD-ImageNet are evaluated using thesettings described in \Cref{Tab:DNN_acuracy}, i.e., each of them is assessed for attack detection to protect victim models inferring on both CIFAR-10 and ImageNet datasets. The experimental results are displayed in \Cref{Fig:HoDS_generalisation}. The results indicate a high generalization capacity across datasets. For instance, GWAD-CIFAR10 classifies the CIFAR-10 validation feature set with an accuracy of $98.22\%$ and also achieves a high classification accuracy of $97.32\%$ on the ImageNet validation set. Similarly, GWAD-ImageNet, achieves $98.94\%$ classification accuracy on ImageNet and achieves $86.36\%$ classification accuracy on CIFAR-10.  It is worth noting that this accuracy corresponds to the multi-class classification including not only detecting but also \textbf{recognizing the attack type}. \textit{The detection rate remains $100\%$ on these benchmarks}, including in the generalization setting.

The confusion matrix reveals a pattern in GWAD's attack recognition performance, with misclassification occurring between queries from certain attacks, such as Sign-Opt and SimBA. This can be attributed to these attacks sharing common primary components in their \ds distributions, as illustrated earlier in  \Cref{Fig:feature_distribution}. 
For example, SimBA and Sign-Flip attacks exhibit a common primary \ds component at $0$ and $-1$.

In the remainder of the paper, we carry on the experiments with GWAD-CIFAR10, and we refer to it as GWAD.  
\begin{table}[t]
\centering
  \begin{tabular}{c|lccc}
	\toprule			
	Familiarity & Attack &  Blacklight & PIHA & GWAD     \\
	\midrule
    \multirow{3}{*}{Known} & {BA}        & 23.96\% & 38.08\% &  99.98\% \\
                           & {HSJA}      & 97.86\% & 98.75\% & 100.00\% \\
                           & {NES}       & 99.96\% & 94.66\% & 100.00\% \\
        \midrule
    \multirow{2}{*}{Unknown} & {QEBA}    & 96.51\% & 96.78\% & 100.00\% \\
                             & {Surfree} & 98.77\% & 70.96\% & 100.00\% \\
	\bottomrule                        
  \end{tabular}
  \caption{GWAD sensitivity is compared to benchmark stateful detections; Blacklight\cite{BLACKLIGHT} and PIHA\cite{PIHA} over the sequences of 100K attack queries against MobileNet-v2 trained on CIFAR-10.}
  \label{Tab:sensitivity_compare}  
\end{table}
\subsection{Sensitivity Evaluation}
\label{sec:sensitivity_evaluation}
We evaluate GWAD's attack detection performance by conducting six query-based black-box attacks against EfficientNet and MobileNet-V2 for ImageNet and CIFAR-10, respectively. 
In every instance, each method performs an untargeted attack until the query budget is reached. GWAD monitors the sequence of the attack queries and extract HoDS at $500$ random checkpoints per attack. 
The results are shown in \Cref{Tab:gwad_sensitivity}, which depicts both attack detection, which is the main objective, and the attack recognition to illustrate GWAD's capacity to distinguish the different attacks' patterns in the \ds space.

\begin{table}[t]
  \begin{tabular}{lccc}
	\toprule			
	Sequence &  Blacklight & PIHA & GWAD     \\
	\midrule
    {CIFAR-10}      & 00.00\% & 00.00\% & 00.00\% \\
    {Tiny-ImageNet} & 00.00\% & 00.01\% & 00.00\% \\
    {ImageNet}      & 00.16\% & 00.14\% & 00.03\% \\
    {FLIR ADAS(RGB)} & 02.17\% & 01.43\% & 00.00\% \\
    {BIRDSAI(Gray)}  & 16.92\% &  --    & 01.29\% \\
	\bottomrule                        
  \end{tabular}
  \caption{False positive comparison on the sequence of queries with benign examples which represent practical use cases.} 
  \label{Tab:false_positive}  
\end{table}
GWAD correctly detects over $99.42\%$ attack queries across all attack methods. Similar to the observation in \Cref{sec:generalisation}, the attack recognition is is affected in some cases due to confusion between attacks with similar \ds distribution. 
We also compare the performance of GWAD with state-of-the-art stateful detection approaches, namely, Blacklight~\cite{BLACKLIGHT} and PIHA~\cite{PIHA}. \Cref{Tab:sensitivity_compare} shows the comparative attack detection performance over the $100K$ attack queries on CIFAR-10/MobileNet-V2. The proposed GWAD achieves near perfect detection rates while other benchmarks largely fail in particular of BA attack. We note that QEBA\cite{QEBA} and Surfree\cite{SURFREE} are unknown to GWAD.
\subsection{Specificity Evaluation}
In this experiment we evaluate GWAD over the sequence of queries with various benign examples and measure the false positive rate (FPR) comparatively with related work. Five publicly available datasets which represent specific real world scenarios are used as presented in \Cref{Tab:false_positive}.
\\ 
\noindent \textbf{Use case: Multi-Class images}
We first conduct tests with image classification tasks, including Tiny-ImageNet, CIFAR-10, and ImageNet, which generally exhibit low similarity between images. Each task is shuffled, and we randomly generate 10K images for monitoring. GWAD does not react to the sequence of images from CIFAR-10 and Tiny-ImageNet, while only $0.03\%$ of ImageNet images in the sequence trigger a false alarm from GWAD.\\
\noindent \textbf{Use case : Urban Scene images }
FLIR ADAS dataset \cite{FLIR_ADAS_V2} is specifically designed for autonomous vehicles and consists of urban scene images in both RBG and IR formats. Since images are captured continuously as a vehicle moves through urban areas, FLIR ADAS dataset includes series of images which have relatively high similarity. GWAD monitors the test sequence and achieves $0\%$ FPR. \\
\noindent \textbf{Use case : Bird's eye images }
BIRDSAI dataset \cite{BIRDSAI} is an infra-red image dataset specifically designed for surveillance systems with the Aerial Intelligence, and due to the large perspective, the images show high similarity. The dataset provides $\sim40K$ night-time images of animals and humans captured from airborne vehicles, and only $1.29\%$ of images trigger false alarms by GWAD. 
We note that PIHA \cite{PIHA} is not applicable to BIRDSAI (gray scale) as hue is 0.\\
\begin{figure}[t]
  \centering
  \includegraphics[width=\linewidth]{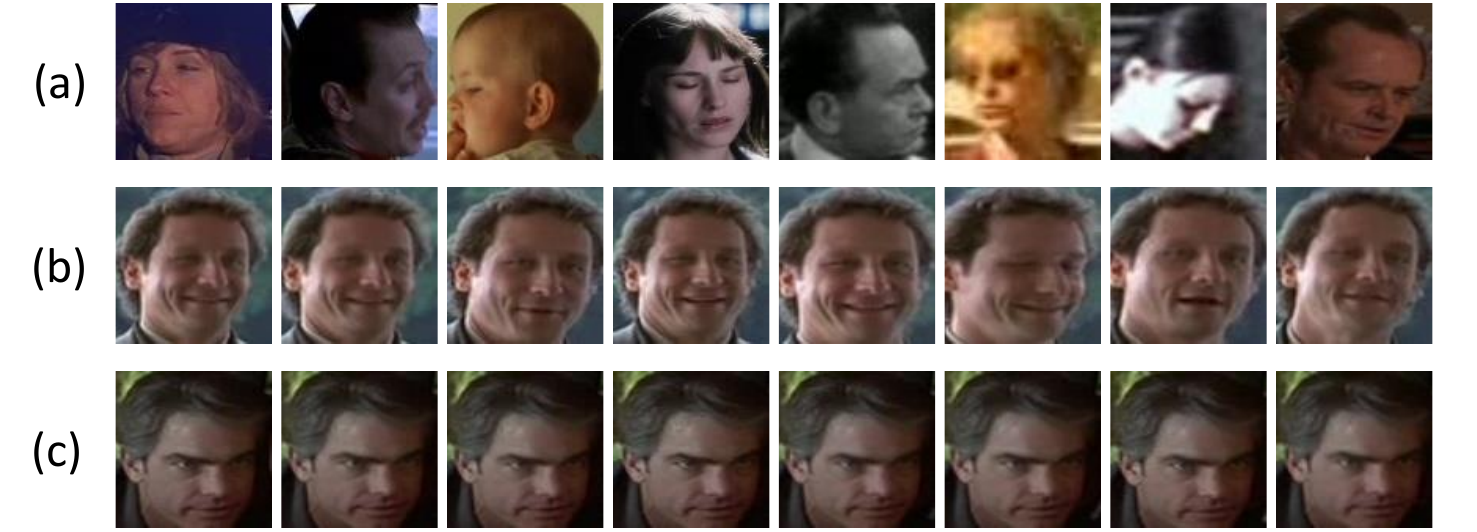}
  \captionof{figure}{Examples of Hollywood Head images: (a) low similarity images; (b) medium similarity images; (c) high similarity images} 
  \label{Fig:hollywood_head} 
\end{figure}
\begin{table}[t]
  \centering  
    \begin{tabular}{lccc}
	\toprule				
	\multirow{2}{*}{Image Similarity}  & \multicolumn{3}{c}{FPR}    \\
        \cmidrule{2-4}
  	    &   Blacklight & PIHA   & GWAD \\
	\midrule
    {Low }      &    02.97\% & 04.55\% & 00.00\% \\	
    {Medium }   &    00.71\% & 02.25\% & 00.00\% \\	
    {High }     &    25.47\% & 26.19\% & 17.99\% \\	
	\bottomrule                        
   \end{tabular}
   \captionof{table}{False Positive comparison: GWAD, Blacklight\cite{BLACKLIGHT} and PIHA\cite{PIHA} monitor the sequence of similar images from Hollywood head dataset}
   \label{Tab:blacklight_benchmark} 
\end{table} 
\noindent \textbf{Use case: Successive similar images:}
We evaluate our approach using the Hollywood Heads dataset \cite{HOLLYWOOD_HEADS}, which provides a sequence of queries containing single-class examples. In addition to a standard split with shuffled examples, the dataset includes a special subset of human head images annotated from sequential movie frames, as illustrated in \Cref{Fig:hollywood_head}. This subset exhibits high similarity between adjacent images. As shown in \Cref{Tab:blacklight_benchmark}, GWAD demonstrates a lower False Positive Rate (FPR) than Blacklight when handling sequences of highly similar, single-class images. Specifically, GWAD achieves an FPR of $17.99\%$, compared to Blacklight $25.47\%$ and PIHA $26.19\%$.

  
\section{Adaptive Attacks}
\label{Sec:adaptive_attack}
In this section, we evaluate our approach under strong adversaries to investigate the limits of GWAD.  
An important requirement for defense methods is their ability to withstand adaptive attacks that have full knowledge of the defense strategy, including the \ds-based approach and GWAD implementation. With this in mind, we consider two types of adaptive attacks: \textbf{i)} Benign example injection, and \textbf{ii)} OARS \cite{OARS}, which is a SOTA attack that has been crafted to bypass stateful defenses.
We provide results for a third adaptive attack in the Supplementary Material (Section 6) consisting of a moving target attack by dynamically varying parameters of the random distribution in the $0^{th}$-order optimization. 
\subsection{Irregular Batch: Benign Example Injection} 
\label{sec:adaptive:irregular_batch}
\begin{figure}[t]
\centering
 \includegraphics[width=0.49\textwidth]{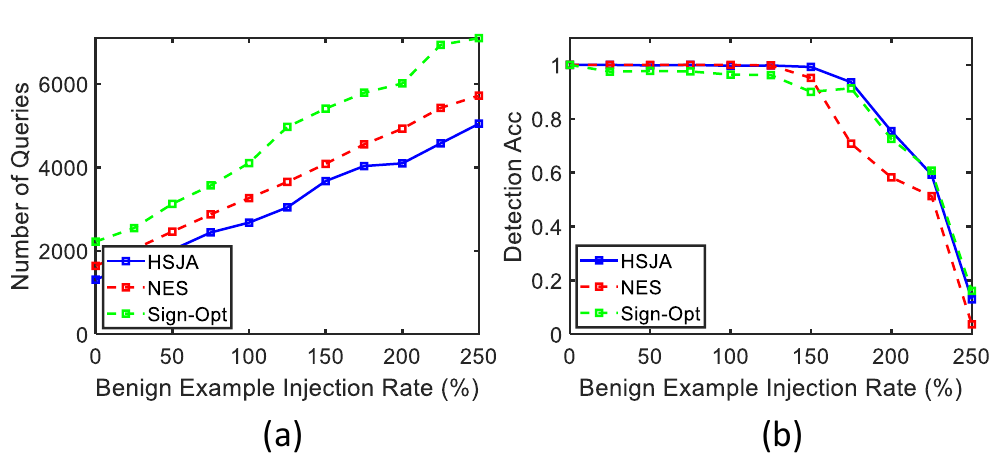}
\caption{(a) shows the number of queries required to achieve successful attacks as benign example injection rate grows for HSJA, NES and Sign-Opt; (b) shows GWAD attack detection accuracy.} 
\label{Fig:adaptive_benign}
\end{figure}

An attacker with prior knowledge of our approach may inject benign examples between malicious (attack) queries to influence the \ds distribution and evade detection.  
To evaluate the robustness of our approach under this scenario, we define the rate of benign example injections ${r}_{b}$ as: $\textbf{r}_{b}= \frac{\#benign}{\#attack-queries}$. We assume that $\textbf{r}_{b}$ is controlled by the adversary and we vary this rate incrementally to explore the limits of our defense. We implement adaptive attacks while varying $\textbf{r}_{b}$ until the attack totally evades detection. \Cref{Fig:adaptive_benign}-(b) displays the evolution of GWAD attack detection accuracy in this setting. Interestingly, GWAD shows a persistent perfect attack detection performance until $\textbf{r}_{b} \leq 150\%$, and gradually degraded after. The detection remains up to $80\%$ for  $r_b \simeq 200\%$. This means that the attacker needs over $2.5 \times$ the overall number of queries to evade detection with $20\%$ success. 

\noindent\textbf{Economics of the attack :} \cref{Fig:adaptive_benign}-(a) shows the average number of queries to achieve successful attacks. The required number of queries to generate successful attacks increases linearly with $\textbf{r}_{b}$. For example, Sign-Opt requires  $2221$ queries at $\textbf{r}_{b} = 0\%$, and it is increased to $7095$ queries at $\textbf{r}_{b} = 250\%$.  Note that an attack being successfully generated in \cref{Fig:adaptive_benign}-(a) does not mean that it evades detection.  
%
\subsection{OARS Adaptive Attack}
Feng et al.~\cite{OARS} proposed a new attack, Oracle-guided Adaptive Rejection Sampling (OARS), to evade stateful defenses~\cite{BLACKLIGHT, STATEFUL_DETECT, PIHA}. OARS identifies the direction of perturbation (through random noise), which sidesteps adaptive stateful detection methods, before injecting the modified example into the target model.
To evaluate the robustness of GWAD against OARS, we conduct experiments using the original OARS source code~\cite{OARS_CODE}. The OARS-NES attack is applied on ResNet-20 with the CIFAR-10 test split, using the Blacklight configuration as specified in the original work. We vary the strength of the random search direction, $\sigma^{2}$, to control and explore its impact on the defenses. 

As shown in \Cref{Tab:oars_gwad}, OARS-NES attack successfully evades Blacklight~\cite{BLACKLIGHT} and PIHA~\cite{PIHA} with the search strength of $\sigma^{2} = 0.05$ achieving $97.5\%$ and $84.5\%$ ASR, respectively. In contrast, OARS-NES fails to bypass GWAD across all test settings. Notably, GWAD perfectly defends against OARS-NES attacks, and its detection rate remains unaffected by changes in $\sigma^{2}$. This is because OARS explores directions in the input space, aiding its evasion of stateful defenses. However, GWAD focuses on analyzing the query update patterns rather than the input space itself.

We also compare the ASR of various OARS adaptive attacks over GWAD comparatively with Blacklight and PIHA, and present the results in \Cref{Tab:oars_asr}. While GWAD shows high robustness against the adaptive attacks, stateful defenses are largely evaded by the attacks.
\begin{table}[t] 
  \centering
  \begin{tabular}{lccc}
	\toprule			
       $\sigma^{2}$ & Blacklight & PIHA & GWAD      \\
	\midrule
        0.001 & 0.0\% / 99.2\% & 0.0\% / 90.8\% & 0.0\% / 100.0\%\\
        0.005 & 1.5\% / 95.8\% & 21.0\% / 63.2\% & 0.0\% / 100.0\%\\
        0.010 & 94.5\% / 3.9\% & 40.4\% / 53.3\% & 0.0\% / 100.0\%\\
        0.050 & 97.5\% / 0.0\% & 84.5\% / 31.7\% & 0.0\% / 100.0\%\\
	\bottomrule                        
  \end{tabular}
  \caption{OARS-NES~\cite{OARS} adaptive attack performances over Blacklight\cite{BLACKLIGHT}, PIHA\cite{PIHA}, and GWAD. Results are presented in (ASR($\downarrow$)/Detection Rate($\uparrow$)) format.}
  \label{Tab:oars_gwad}  
\end{table}
\begin{table}[t]  
  \centering
  \begin{tabular}{lccc}
	\toprule			
        {Attack} & {Blacklight} & {PIHA} & {GWAD}  \\        
	\midrule
        OARS-NES      &  $98\%$ &  $82\%$ & $00\%$\\
        OARS-HSJA     &  $75\%$ &  $71\%$ & $00\%$\\
        OARS-QEBA     &  $98\%$ &  $95\%$ & $00\%$\\
        OARS-Square   &  $96\%$ & $100\%$ & $05\%$\\
	\bottomrule                        
  \end{tabular}
  \caption{Attack Success Rate (ASR) of various OARS adaptive attacks over Blacklight~\cite{BLACKLIGHT}, PIHA~\cite{PIHA}, and GWAD. The lower ASR is the better defence.} 
  \label{Tab:oars_asr}  
\end{table}
%

\section{GWAD$^+$: Enhancement with benign prescreening}\label{sec:enhancement}

In the previous section, we found that GWAD was relatively vulnerable to benign example injection, even though it results in significantly higher cost for an attack to converge. This section investigates ways to enhance GWAD further. 
Since GWAD and stateful defenses operate in different spaces, we propose to enhance GWAD with a stateful pre-screening. 
Specifically, we use a lightweight stateful pre-processing step that we name "\textit{Screener}" to exclude non-suspicious queries using an input space analysis. 
The Screener transforms a query into $1 \times 128$ vector $e_{0}$ through the following steps: \textbf{i)} rescale to $32 \times 32$ image space, \textbf{ii)} convert to binary image with Canny edge detection~\cite{CANNY_EDGE}, \textbf{iii)} compress each row to 4-Byte by representing binary pixels to bits. Finally the transformed query is pushed into $n$-depth FIFO ($n$ previous queries). The \textbf{similarity} between input query $e_{0}$ and queries $e_{i}$ in the FIFO is calculated as: 
\begin{equation}
  m_{i}\ = \left.
  \begin{cases}
    \;\;\,1,   \;\; \text{if} \; \frac{\hat{S}(e_{0} \oplus e_{i})}{\hat{S}(e_{0}) + \hat{S}(e_{i})} < \theta \\
    \;\;\,0,   \;\; \text{otherwise.}      
  \end{cases}
  \right.
\label{Eq:Ch3_SQBA_H(x)_Boolean}
\end{equation}
where $i = \{1,2,.., n\}$, $\oplus$ is bit-wise xor, $\hat{S}$ is the number of bit 1, and $\theta$ is an empirically defined threshold. 
\begin{figure}[t]
\centering
 \includegraphics[width=\columnwidth]{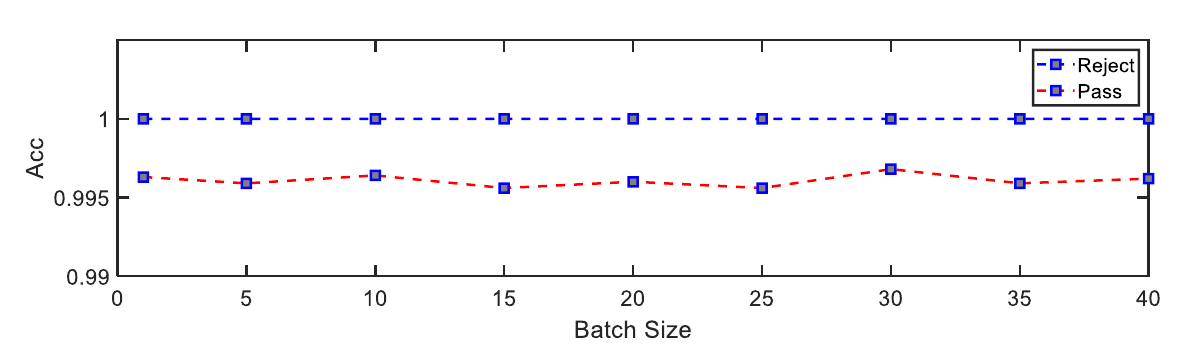}
\caption{Screener pre-process performance over irregular batch attacks based on HSJA with various batch sizes.}
\label{Fig:pre_process_performance}
\end{figure}
The performance of the screener is shown in \cref{Fig:pre_process_performance}. Irregular batch attacks based on HSJA conduct 100 attacks on CIFAR-10 Mobile-Net with various batch sizes. Regardless of the batch sizes, the benign queries are perfectly screened out at $100\%$ accuracy, and over $99.5\%$ of attack queries are fed into GWAD for the precise attack detection. Although FIFO depth is set to $n = 100$ in this experiment, we tested with different depth values without any significant impact on the performance. Further details can be found in the supplementary Section 5. 
%
\begin{table}[t!]
\centering
\begin{tabular}{lcccc}
\hline
\multirow{2}{*}{Attack} & \multicolumn{2}{c}{ImageNet} & \multicolumn{2}{c}{CIFAR-10} \\ 
\cline{2-5} 
    & GWAD & $\text{GWAD}^{+}$   & GWAD & $\text{GWAD}^{+}$   \\ 
\hline
HSJA      & 00.00\%  & 99.83\% & 00.12\%      & 99.24\%          \\
NES       & 00.00\%  & 99.95\% & 00.00\%      & 99.93\%          \\
Sign-Opt  & 00.00\%  & 99.50\% & 00.01\%      & 99.50\%          \\
Sign-Flip & 00.01\%  & 99.72\% & 00.01\%      & 99.66\%          \\
\hline
\end{tabular}
\caption{Enhanced performance of $\text{GWAD}^{+}$ (Screener + GWAD) over irregular batch attack with $\textbf{r}_{b} = 1000\%$ (100 benign for 10 malicious queries) against EfficientNet trained on ImageNet and MobileNe-v2 trained on CIFAR-10.}
\label{Tab:gwad_screen_sensitivity}
\end{table}

We refer to the Screener-enhanced GWAD as $\text{GWAD}^{+}$ and conduct an experiment to evaluate the enhanced performance over the irregular batch attacks. The test setting used in \cref{sec:sensitivity_evaluation} is adapted in this experiment, and the attacks are performed with $\textbf{r}_{b} = 1000\%$, i.e., injecting benign queries that are $10\times$ the number of actual attack queries to strongly obfuscate the attack. \cref{Tab:gwad_screen_sensitivity} compares the detection rates of GWAD and $\text{GWAD}^{+}$. While GWAD is completely evaded by such adaptive attacks, $\text{GWAD}^{+}$ shows highly robust performance achieving over $99.2\%$ detection rate.

\section{Related Work}
\label{Sec:related_works}
The problem of defending against adversarial attacks is a highly active research area. However, while query-based black-box attacks are the most practically relevant attack scenarios,  there are relatively few approaches that specifically target these threat models. 
Chen et al. \cite{STATEFUL_DETECT} introduced Stateful Detection (SD), which seeks to identify adversarial example generation in black-box adversarial attacks. SD monitors sequences of queries, reducing each input example to a lower-dimensional representation via a pre-trained neural network feature extractor. It then calculates pairwise $l_2$ similarities between a new query and its $k$-nearest neighbors among previous queries to identify potential attack queries. However, SD is resource-intensive and has lower detection accuracy compared to Blacklight \cite{BLACKLIGHT}. 
Li et al. \cite{BLACKLIGHT} (Blacklight) introduces a quantization process that maps a new query to a compact vector space, generating dedicated $S \times 32$ byte hash values. Blacklight then compares these hash values to those stored in a dictionary along with their corresponding query IDs. It uses matched hash values to assess similarity between queries, adding unmatched hashes to the dictionary for future reference. When a query ID in the dictionary accumulates matched hash values beyond a set threshold, Blacklight flags the query as malicious.
Another similar stateful detection approach called Perceptual Image HAshing (PIHA) has been proposed for adversarial query detection \cite{PIHA}. Specifically, PIHA uses statistical methods to analyze the input history, identifying anomalies in the distribution of queries.

However, we found that these defenses suffer from a high FPR under high similarity benchmarks. Moreover, stateful defenses have shown vulnerable to OARS \cite{OARS}. This attack uses initial query patterns to infer key properties about an SD defense; and, craft subsequent query patterns to evade the defense while making progress towards finding adversarial inputs. Using the \ds, we show that, unlike existing defenses, GWAD is robust against OARS, along with a negligible FPR.

\section{Concluding remarks}
\label{Sec:conclusion}
In this paper, we propose a novel approach for detecting black-box adversarial attacks that overcomes the vlnerabilities of stateful defenses. Specifically, we introduce Gradient Watch Adversarial Detection (GWAD), a defense framework designed to detect query-based black-box attack attempts. We introduce a new concept, \ds, as a means of analyzing query behavior. The main insight of the proposed approach is that \ds allows for a distinctive differentiation in the input update space. Our experiments demonstrate that GWAD achieves high detection performance against six state-of-the-art black-box attack methods, along with an adaptive attack (OARS) that evades stateful detection approaches. Besides, our results show a model and dataset-agnostic property of our approach. In fact, the attack detector does not to be trained on the same data distribution of the protected model. 
One limitation of GWAD is that it can be evaded with excessive benign data injection within the attack queries. With the screening enhancement, we demonstrate that GWAD$^+$ is robust against benign injection in irregular batches.
We believe that this work contributes a new perspective to the adversarial attack problem in black-box settings, advancing robust defenses under practical threat models towards trustworthy machine learning.

{
    \small
    \bibliographystyle{ieeenat_fullname}
    \bibliography{bib_db}
}


\begin{center}
\textbf{\Large Supplementary Materials}
\end{center}

\setcounter{section}{0}

\section{Concentration of Measure}
\label{sec:supp::concentration}
\ds distributions extracted from the sequence of queries from query-based black-box adversarial attacks exhibit unique and distinguishable patterns. To explain this observation, we take an approach that adopts the concentration of measure phenomena in high-dimensional spaces. The properties of high-dimensional spaces often defy the rules based on low-dimensional spaces we are familiar with\cite{BLUM_DATA_SCIENCE}, and there is a set of less appreciate phenomena, especially in data analysis, called the concentration of measure. The concentration of measure phenomena are non-trivial observations and properties of the large number of random variables\cite{CONCENTRATION_OF_MEASURE}. 
Two important and useful geometric metrics of random distributions are the length of a random vector $u_{i} \in \mathbb{R}^{d}$, and the angle of two random vectors $u_{i}, u_{j} \in \mathbb{R}^{d}$. In high-dimensional spaces, however, the concentration of measure states these metrics are almost concentrated to a single value in the sense of the measure. \\
\subsection{Concentration of Length}
\label{sec:supp:concentration:length}
Let $d$-dimensional vector $u_{i} \in \mathbb{R}^{d}$ be a random vector that is sampled from a random distribution $\mathcal{N}(\mu, \sigma^{2})$. The law of large numbers states that $\frac{1}{d}\sum_{k=1}^{d} u_{i}^{k}$ is almost surely $\mu$ as $d \rightarrow \infty$~\cite{EVANS_BOOK}, where $u_{i}^{k}$ is the $k$'th element of $u_{i}$. If $\mu$ and $\sigma^{2}$ is fixed, the length of any vector $u_{i}$ is expected to converge toward a common value. Especially when $\mu = 0$, the length converges to $\sigma^{2} \sqrt{d}$.

Suppose a random distribution $\mathcal{N}(\mu, \sigma^{2})$ has $\mu = 0$, and the random fluctuation is limited with the unit variance $\sigma^{2} = 1$. The expected length of $u_{i}$ is denoted as\cite{BANDEIRA_LECTURE_NOTES},
\begin{equation}
  E \left[ \| u_{i} \|_{2}^{2} \right] = E \left[ \sum_{k=1}^{d} |u_{i}^{k}|^{2} \right] = \sum_{k=1}^{d} E \left[ |u_{i}^{k}|^{2} \right] = d 
\label{Eq:length_concentration_0}
\end{equation}
where $E [ u_{i}^{k} ]$ is the expected value of the random variable $u_{i}^{k}$. Therefore, the Euclidean length of $u_{i}$ is expected to be approximately $\sqrt{d}$. \\
\begin{figure}[t]
  \includegraphics[width=0.48\textwidth]{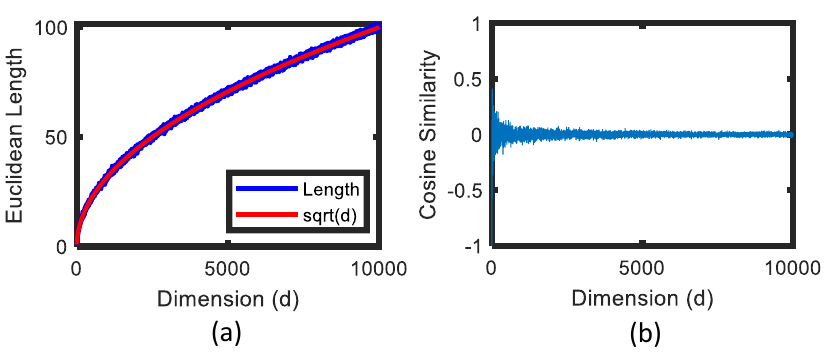}
  \captionof{figure}{(a) Euclidean length of random vectors with the unit variance $\sigma^{2} = 1$ follows $\sqrt{d}$ as the dimension $d$ grows; (b) Cosine Similarity of two random vectors converges toward 0 ($90^{\circ})$ as the dimension $d$ grows.}
  \label{Fig:concent_measure}
\end{figure}
\subsection{Concentration of Angle}
\label{sec:supp:concentration:angle}
Let two $d$-dimensional vectors $u_{i}$, $u_{j} \in \mathbb{R}^{d}$ be independent random vectors with Rademacher variables as $u_{i}^{k} \in \{-1, 1\}$. The angle between two vectors is denoted as $\angle u_{i}u_{j}$, and the cosine similarity is denoted as $\cos\angle u_{i}u_{j} = \frac{ u_{i} \cdot u_{j} }{ \|u_{i}\|_{2} \|u_{j}\|_{2} }$, where $u_{i} \cdot u_{j} = \sum_{k=1}^{d}u_{i}^{k}u_{j}^{k}$ is the sum of independent random variables, hence $E[u_{i} \cdot u_{j}] = \sum_{k=1}^{d} E[u_{i}^{k}u_{j}^{k}] = 0$. Therefore, Hoeffding's inequality\cite{HOEFFDING_INEQUALITY} can be applied for any given $t > 0$ as 
\begin{equation}
	P \left( | u_{i} \cdot u_{j} | \geq t \right) = P \left( \frac{ | u_{i} \cdot u_{j} | }{ d } \geq \frac{t}{d} \right) \leq 2 e^{\left(-\frac{t^{2}}{2d}\right)}
\label{Eq:angle_concentration_0}
\end{equation} 
where $P$ is the probability. Now replace $t$ with $\sqrt{2d\log{d}}$, then the inequality in \Cref{Eq:angle_concentration_0} is rewritten as
\begin{equation}
P \left( \frac{ | u_{i} \cdot u_{j} | }{ d } \geq \sqrt{ \frac{2\log{d}}{d} } \right) \leq 2 e^{\left(-\log{d} \right)}
\label{Eq:angle_concentration_1}
\end{equation} 
From \Cref{Eq:length_concentration_0}, $\|u_{i}\|_{2} \|u_{j}\|_{2}$ is surely $d$. Therefore, \Cref{Eq:angle_concentration_1} becomes as follows \cite{BANDEIRA_LECTURE_NOTES},
\begin{equation}
P \left( | \cos\angle u_{i}u_{j} | \geq \sqrt{ \frac{2\log{d}}{d} } \right) \leq \frac{2}{d}
\label{Eq:angle_concentration_2}
\end{equation} 
As stated in \Cref{Eq:angle_concentration_2}, the angles of two independent vectors, sampled from $\mathcal{N}(0, \sigma^{2})$, highly likely become narrowly distributed around the mean $\angle u_{i}u_{j} = \pi/2$, with a variance that converges towards zero\cite{CURSE_OF_DIM} as the dimension grows.

The concentration of measure phenomena are empirically proven as illustrated in \Cref{Fig:concent_measure}.
\section{Delta Similarity of Attack methods}
\label{Sec:supp:ds_analysis}
\begin{figure}[t]
  \includegraphics[width=0.48\textwidth]{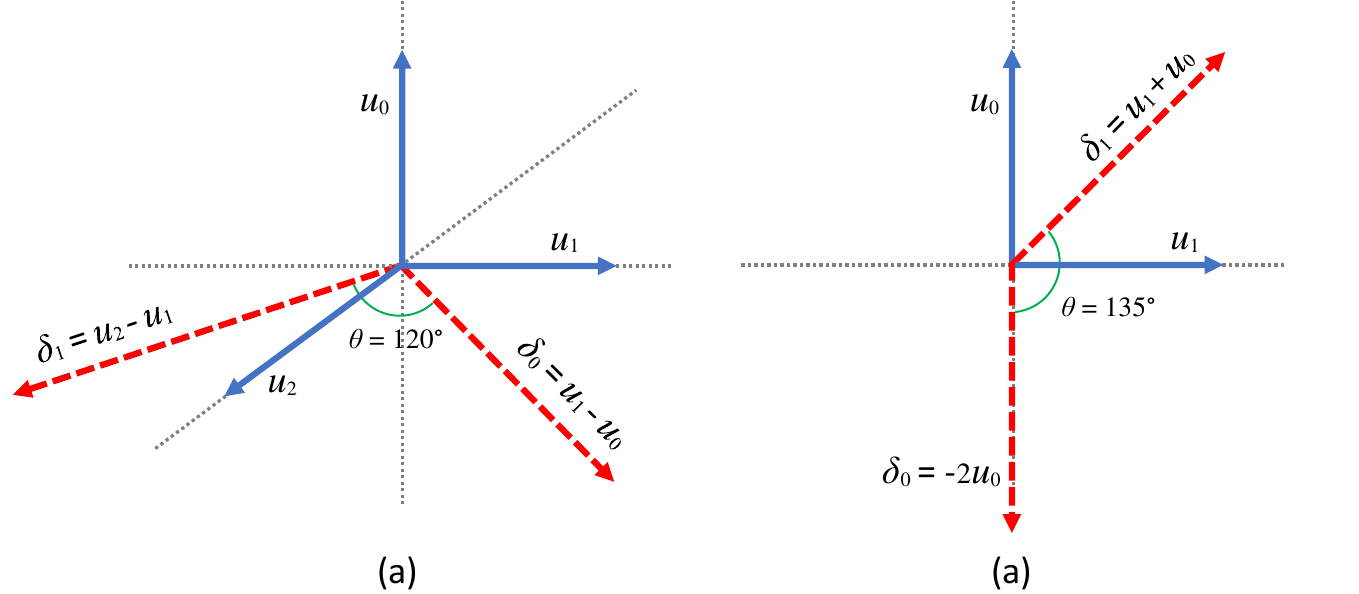}
  \captionof{figure}{Delta Similarity (\ds) of Query-based Black-box Attacks. (a) \ds of HSJA zeroth order optimisation; (b) \ds of NES zeroth order optimistaion.}
  \label{Fig:ds_of_attack}
\end{figure}

We provide \ds analysis of HSJA\cite{HSJA} and NES\cite{NES} attack methods. These two attack methods represent hard-label and soft-label based attack strategies respectively.\\  
\noindent\textbf{HSJA :}
HSJA\cite{HSJA} estimates the direction of gradient via the Monte Carlo algorithm as:
\begin{equation}
	\nabla \mathcal{H}(\tilde{x}_{t}) \approx \frac{1}{n} \sum_{j=1}^{n}\mathcal{H}(\tilde{x}_{t} + \epsilon u_{t}^{j}) u_{t}^{j}
\label{Eq:HSJA_zeroth_order}
\end{equation}
where $u_{t}^{j} \in \mathbb{R}^{d}$ is i.i.d. random vector. The sequence of queries in \Cref{Eq:HSJA_zeroth_order} is $\{\tilde{x}_{t} + \epsilon u_{t}^{j},\;\tilde{x}_{t} + \epsilon u_{t}^{j+1}, \;... \}$. Therefore, $\delta$ becomes a scaled subtraction of two random vectors as $\delta^{i}_{t} = \epsilon(u_{t}^{j+1} - u_{t}^{j})$ and $\delta^{i+1}_{t} = \epsilon(u_{t}^{j+2} - u_{t}^{j+1})$. Three random vectors $u_{t}^{j}, u_{t}^{j+1}$ and $u_{t}^{j+2}$ are surely orthogonal and have the same length. Hence $\mathcal{DS} = -0.5(120^{\circ})$ as illustrated in \Cref{Fig:ds_of_attack} (a).\\
\noindent\textbf{NES :}
Ilyas et al. adopted NES\cite{NES} to estimate the gradient. They use search distribution of random Gaussian noise around the intermediate adversarial example in every iteration such as:
\begin{equation}
	\nabla F(\tilde{x}_{t}) \approx \frac{1}{2 \epsilon n} \sum_{j=1}^{n} \big( F(\tilde{x}_{t} + \epsilon u_{t}^{j}) - F(\tilde{x}_{t} - \epsilon u_{t}^{j}) \big)u_{t}^{j} 
\label{Eq:NES_zeroth_order}
\end{equation}
The sequence of queries in \Cref{Eq:NES_zeroth_order} is $\{ \tilde{x}_{t} + \epsilon u_{t}^{j},\; \tilde{x}_{t} - \epsilon u_{t}^{j},\; \tilde{x}_{t} + \epsilon u_{t}^{j+1},\; ...\}$. In this sequence, $\delta^{i}_{t} = -2\epsilon u_{t}^{j}$, and $\delta^{i+1}_{t} = \epsilon(u_{t}^{j+1} + u_{t}^{j})$. Therefore \ds becomes $-0.7071(135^{\circ})$. 
\ds of NES zeroth order optimisation is illustrated in \Cref{Fig:ds_of_attack} (b).
\section{Screener Pre-Processing}
\begin{figure}[t]
\begin{minipage}[t]{0.48\textwidth}
  \includegraphics[width=\textwidth]{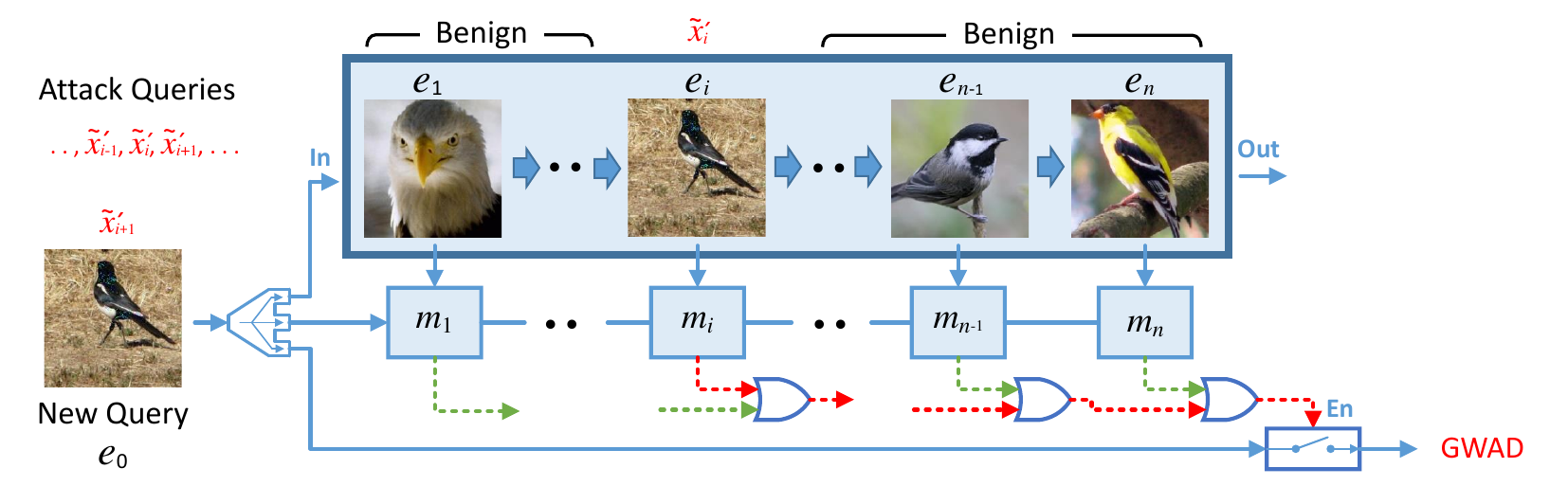}
  \captionof{figure}{Conceptual diagram of GWAD Screener pre-process. Screener screens out un-suspecious examples deliberately injected by the irregular batch attacks.}
  \label{Fig:pre_process}
\end{minipage}
\vfill
\begin{minipage}[t]{0.48\textwidth}
  \includegraphics[width=\textwidth]{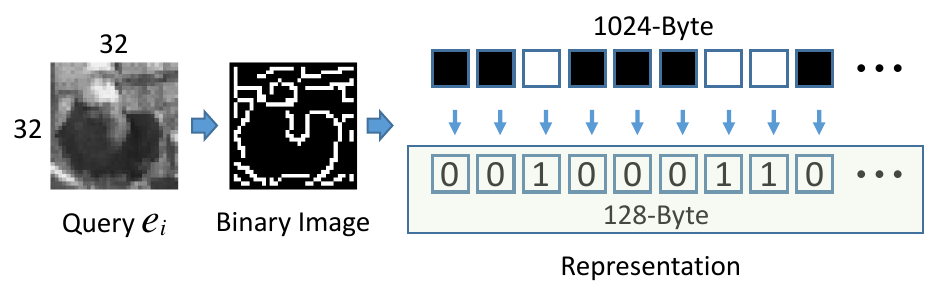}
  \captionof{figure}{GWAD Screener pre-process 128-Byte representation of an example}
\label{Fig:pre_process_feat}
\end{minipage}
\end{figure}

GWAD screener pre-process is a cost-effective and light-weight stateful detection to screen out un-suspicious queries which are deliberately injected by the irregular batch attacks. The screener represents an image $x$ with the 128-Byte vector. Image $x$ first converted to $32 \times 32$ grayscale image, then transformed into a binary image by Canny edge detection algorithm \cite{CANNY_EDGE}. Each pixel of the binary image is represented with bit 0 or 1, and finally forms a 128-Byte vector. The similarity between two images is scored by the rate of mismatching pixels which is simply calculated with the bit-wise xor of represented vectors and the number of bit 1 in the result, as described in Equation 5 of the main paper. \Cref{Fig:pre_process} illustrates GWAD screener pre-process and \Cref{Fig:pre_process_feat} depicts the 128-Byte representation. \\
\textbf{\textit{n}-Channel Attack :} $n$-Channel attack setting denotes where an adversary attacks multiple images in parallel. To cope with the $n$-Channel attack, screener is extended with a channel-aware detection mechanism that assigns a channel ID (CID) to each query, enabling independent tracking of different query sequences:
\textbf{i)} Queries are dynamically assigned to distinct channels based on their similarity to prior inputs.
\textbf{ii)} Each query’s $\mathcal{DS}$ score is computed within its respective channel, preventing adversarial alternation.
\textbf{iii)} Screener efficiently maintains separate state information for each attack sequence, ensuring that alternating queries do not evade detection. 
Our preliminary results confirm the effectiveness of this strategy, achieving over $99\%$ accuracy in detecting a 2-Channel attack 
\begin{table}[t] 
  \centering  
    \begin{tabular}{lcccc}
	\toprule				
	\multirow{2}{*}{Attack}  & \multicolumn{4}{c}{Screener FIFO Depth}    \\
        \cmidrule{2-5}
  	        &   100 & 200 & 600 & 1K \\
	\midrule
        HSJA      &  99.50\% &  99.11\% &  99.61\%  &  98.86\%\\
        NES       &  99.92\% &  99.91\% &  99.92\%  &  99.92\%\\	
        Sign-Flip &  99.76\% &  99.68\% &  99.79\%  &  99.78\%\\	
        \bottomrule                        
   \end{tabular}
  \caption{$\text{GWAD}^{+}$ (Screener + GWAD) detection rates with various Screener FIFO depths over irregular batch attack at $\textbf{r}_{b} = 1000\%$ against ResNet-18 trained on CIFAR-10}
  \label{Tab:effect_fifo_depth}   
\end{table}
\section{Attack Classifier and Training Feature Set}
\label{sec:suppl:classifier}
We train a simple neural network to classify attack queries. As shown in \Cref{Tab:DeltaNet_architecture}, the network consists of six full-connected hidden layers with ReLU activations. The classification probabilities are provided by a Log-Softmax output layer.

\begin{table}[t]
\begin{minipage}[t]{0.48\textwidth}
  \centering
    \begin{tabular}[t]{cccrcc}
    \toprule
        Layer  & Layer Type   &  & Node & & Activation \\
    \midrule
        0      & Input        &  & 201  & & ReLU \\
        1      & Linear       &  & 512  & & ReLU \\
        2      & Linear       &  & 512  & & ReLU \\
        3      & Linear       &  & 256  & & ReLU \\
        4      & Linear       &  & 128  & & ReLU \\
        5      & Linear       &  & 64   & & ReLU \\	
        5      & Linear       &  & 7    & & ReLU \\		
        6      & LogSoftmax   &  & 7    & & -    \\
    \bottomrule                        
    \end{tabular}
  \captionof{table}{Architecture of GWAD attack classifier} 
  \label{Tab:DeltaNet_architecture}     
\end{minipage}
\hfill
\end{table}

\noindent\textbf{HoDS Training Feature Set :}
\begin{figure*}[hbt!]
  \centering
   \includegraphics[width=1.0\linewidth]{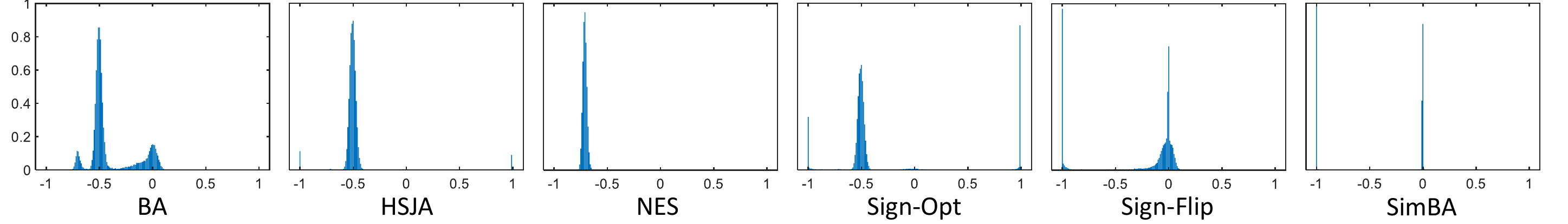} 
   \caption{Mean models of CIFAR-10 HoDS training dataset} 
   \label{Fig:hods_mean_model}
\end{figure*}
We first configure all the adversarial attack methods, listed in \Cref{Tab:list_attack_methods} of the main paper, to carry out their attacks with the query budget $q_{\epsilon} = 5K$ without stopping criteria enabled. To generate a set of HoDS features to train GWAD-CIFAR10, each attack method performs the attack on the pre-trained ResNet-18, with 10 randomly selected examples from the training split. From each attack, 150 HoDS features are extracted at random points in the sequence of queries. Therefore, each attack method generates 1500 HoDS features during its attacks on 10 examples. As a result, 9000 HoDS features are extracted for the six attack classes. 
Another set of HoDS features to train GWAD-ImageNet is also acquired through the same procedure with the pre-trained VGG-16. We present the visualisation of HoDS features in \Cref{Fig:hods_mean_model}. \\
HoDS feature set to represent the benign class in the training are acquired from the normal distributions instead of the real benign queries. We use four normal distributions, and extract 375 HoDS features from each distribution:
\begin{itemize}
\item $\mathcal{N}(0, 0.25)$ and $\mathcal{N}(-0.5, 0.25)$
\item $\mathcal{N}(0, 0.14)$ and $\mathcal{N}(-0.5, 0.14)$
\end{itemize}
The mean and variance of the normal distributions are chosen based on the empirical measure of the $\mathcal{DS}$ distributions. For example, the $\mathcal{DS}$ elements of CIFAR-10 in \Cref{Fig:feature_distribution} of the main paper are randomly distributed with $\mu = -0.49447$, and $\sigma^{2} = 0.14828$. 

\section{Dataset and Attack Methods}
\label{sec:supp:attack_methods}

\subsection{Dataset}
\label{sec:supp:dataset_attack_methods:dataset}
The experiments of query-based black-box attacks on image classification tasks are conducted over two standard image datasets: CIFAR-10 \cite{CIFAR-10} and ImageNet \cite{IMAGENET-1000}. All the examples of these datasets are transformed as instructed by Pytorch torchvision library \cite{IMAGENET_MODELS}. We present further details of image datasets used in the experiments with benign image queries to simulate the practical use cases as follows.\\
\noindent\textbf{Tiny-ImageNet} \cite{TINY_IMAGENET} is composed of 200 classes of images. Images in the dataset are downsized to $64 \times 64$ coloured image space. 
The dataset is widely used for training and testing various machine learning techniques.\\
\noindent\textbf{Hollywood Heads} \cite{HOLLYWOOD_HEADS} is a dataset containing human heads annotated in sequential Hollywood movie frames. As objects are in a single class and extracted from the sequential movie frames, examples tend to exhibit high similarity with neighbouring examples. \\
\noindent\textbf{FLIR ADAS}\cite{FLIR_ADAS_V2} provides thermal and visible band images for the development of automated systems using modern DNN models. The dataset was acquired via camera system mounted on a vehicle, and includes images captured in streets and highways in California, USA.    \\ 
\noindent\textbf{BIRDSAI}\cite{BIRDSAI} is an infra-red image dataset specifically designed for Surveillance system with Aerial Intelligence. The dataset was acquired through a long-wave band thermal camera system, and contains night-time images of animals and humans in Southern Africa. 
\subsection{Attack Methods}
\label{sec:supp:dataset_attack_methods:attack}
\begin{table}[t]
  \centering
  \begin{tabular}{lccc}
    \toprule
    \multirow{2}{*}{Attack methods} & \multicolumn{3}{c}{Query Consumption} \\
    \cmidrule{2-4}
        & Zeroth Opt.     & Linear Search    & Other   \\
    \midrule
    BA \cite{BA}              &  $90.42\%$ & $00.50\%$  & $9.08\%$\\
    HSJA \cite{HSJA}          &  $96.06\%$ & $03.93\%$  & $0.01\%$\\
    SimBA \cite{SIMBA}        & $100.00\%$ & $00.00\%$  & $0.00\%$\\
    Sign-OPT \cite{SIGN_OPT}  &  $71.67\%$ & $28.14\%$  & $0.19\%$\\
    Sign-Flip \cite{SIGN_FLIP}&  $99.80\%$ & $00.19\%$  & $0.01\%$\\
    NES \cite{NES}            & $100.00\%$ & $00.00\%$  & $0.00\%$\\
  \bottomrule
  \end{tabular}
  \caption{Query consumption of SOTA attack processes. Attacks on MobileNet-V2 trained on CIFAR-10 with 5K query budgets }
  \label{Tab:query_consumption}  
\end{table}
We note that all the attack methods used in the experiments are commonly configured to the untargeted setting with $l_{2}$ constraint. In the following, we provide the hyper-parameter settings of the methods used in the experiments.\\
\noindent\textbf{BA}\cite{BA} : Optimisation steps are initialised with 0.01, and updated every 10 iterations with a learning rate $\epsilon = 1.5$. \\
\noindent\textbf{HSJA}\cite{HSJA} : 100 queries are used to find an initial adversarial example. Binary search threshold $\theta$ is set to $0.01 / \sqrt{w\times h\times c}$, where $w\times h\times c$ is the area of input space. \\
\noindent\textbf{SimBA}\cite{SIMBA} : Attack step size $\epsilon$ is set to 0.03 for CIFAR-10, and 0.2 for ImageNet.  \\
\noindent\textbf{Sign-Opt}\cite{SIGN_OPT} : Gradient search learning rates are initialised to 0.001 for CIFAR-10, and 0.05 for ImageNet. Line search learning rates are set to 2 and 0.25 for further optimisation. Finally, convergence threshold is 255 and 5 for CIFAR-10 and ImageNet respectively. \\
\noindent\textbf{NES}\cite{NES} : Number of samples $n$ for the gradient estimate is 50. Learning rate $\eta$ is set to 0.55 for CIFAR-10 and 2.55 for ImageNet. Search variance $\sigma$ is 0.1. \\
\noindent\textbf{Sign-flip}\cite{SIGN_FLIP} : Project step parameter $\alpha$ is initialised to 0.0004 and updatd with a rate of 1.5. Random sign flip step parameter $p$ is initialised to 0 and updated with a step of 0.001. 

Query-based attack methods spend the majority of their attack queries
on zeroth-order optimisations. \Cref{Tab:query_consumption} shows the query consumption profile of six SOTA query-based black-box attack methods based on the settings detailed above.  
\section{Additional Experiments}
\label{sec:supp:additional_benchmark}

\subsection{Moving target attack}
In this section, we consider an attacker that implements a "moving-target" strategy to evade detection, while still generating adversarial examples and querying the victim model. The idea is to manipulate the parameters of the noise distribution from which the attack samples the noise. 

Noise vectors $u_t$ used in zeroth-order optimisation need to be the zero-mean random distributions with a common variance such as $\mathcal{N}(0, 1)$ to guarantee acceptable quality of attacks (QoA). However, one adaptive attack, with full knowledge of the proposed detection scheme, may introduce a variation in the random distributions by varying $\mu$ or $\sigma^{2}$, with no serious consideration of QoA. 
%
\begin{table}[t]
  \centering  
    \begin{tabular}{lccc}
    \toprule			
      \multirow{2}{*}{Variance Bound}  &  \multicolumn{3}{c}{Detection Rate}  \\
      \cmidrule{2-4}
                       & HSJA & NES & Sign-Flip\\
	\midrule
    $0.0\leq \alpha \leq 1.5$  &  100.00\% &  100.00\% & 99.30\%\\
    $0.0\leq \alpha \leq 2.0$  &  100.00\% &  100.00\% & 99.28\%\\
    $0.0\leq \alpha \leq 2.5$  &  100.00\% &   99.98\% & 99.39\%\\ 
    $0.0\leq \alpha \leq 3.0$  &  100.00\% &  100.00\% & 99.42\%\\ 
	\bottomrule                        
    \end{tabular}   
  \captionof{table}{GWAD attack detection performances over the sequence of queries from varying-variance adaptive attacks with HSJA \cite{HSJA}, NES \cite{NES} and Sign-Flip \cite{SIGN_FLIP}}
  \label{Tab:adaptive_variance}    
\end{table}

\noindent\textbf{Varying-variance attack : }
We first conduct attacks with varying variance of random distributions as presented in \Cref{Tab:adaptive_variance}.
In this setting, generated random noises are scaled by a factor $\alpha$, where $0 \leq \alpha \leq \textbf{r}_{\sigma}$. While the attack performance of NES has gradually degraded, GWAD maintains the attack detection performance showing near perfect detection rates against varying variance adaptive attacks, based on HSJA, NES and Sign-Flip method, across all the variations in $\sigma^{2}$.

\begin{figure}[t]
  \centering
  \includegraphics[width=0.47\textwidth]{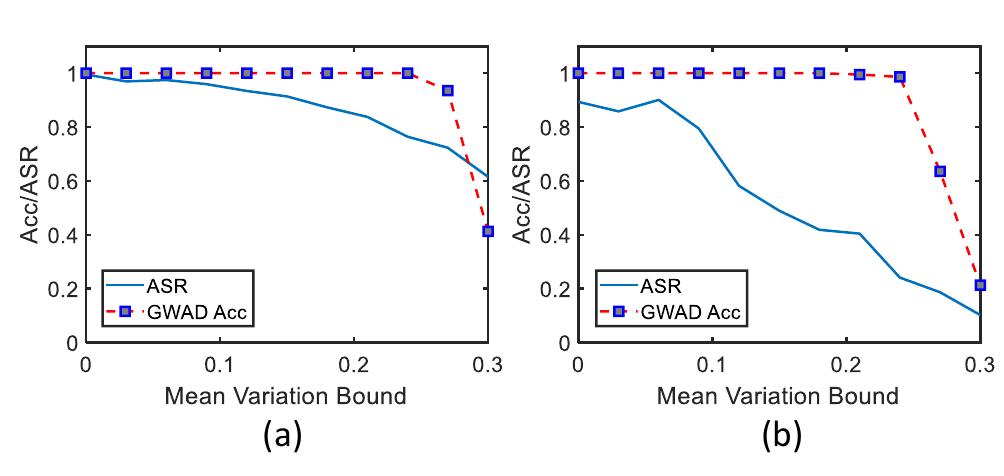}
  \captionof{figure}{ASR of varying-mean adaptive attacks and detection rates of GWAD as the upper bound $\textbf{r}_{\mu}$ of $\mu$ variation in random distributions grows: (a) HSJA; (b) NES}
\label{Fig:vary_random_result}
\end{figure}

\noindent\textbf{Varying-mean attack: }We now consider the scenario where the adversary varies the mean $\mu$ of random vectors. The attack first finds the range of pixel intensities for an input $x$ as $s = \max(x) - \min(x)$, and sets the bounds of variation as a ratio $\textbf{r}_{\mu}$ of $s$. We display the impacts on ASR caused by this adaptive attack setting in \Cref{Fig:vary_random_result}. The ASR of such attacks are gradually decreased as the rate of variation grows. The adaptive attacks with HSJA and NES show only $61.54\%$ and $10.2\%$ ASRs respectively at $\textbf{r}_{\mu}=0.30$. In contrast, GWAD maintains the robustness in binary classification performance until $\textbf{r}_{\mu}=0.24$ achieving $100\%$ detection rates against both attacks. \\ 
\subsection{Other Classifier architectures}
We explore conventional classifier architectures to evaluate the discriminant power of HoDS features. Among these architectures, the classification performances of two non-DNN classifiers, SVM and kNN, are presented in \Cref{Tab:non-dnn}. We note that the kNN classifier is directly applied with the HoDS-Mean of CIFAR-10 depicted in \Cref{Fig:hods_mean_model} at $k=256$.  

\begin{table}[t] 
\centering
\scriptsize
  \begin{tabular}{lcccccc}
	\toprule			
	   & HSJA & NES & SimgBA & Sign-Opt & Sign-Flip & BA     \\
	\midrule
    {SVM}  & 100.0\% & 100.0\% & 100.0\% & 84.1\% & 99.7\% & 94.9\%\\
    {kNN} &  89.4\% & 100.0\% & 100.0\% & 93.7\% & 92.6\% & 85.6\%\\
	\bottomrule                        
  \end{tabular}
  \caption{Non-NN models' classification performance (ImageNet)} 
  \label{Tab:non-dnn}  
\end{table}
%
\subsection{Ablation Studies}
\begin{table}[t] 
  \centering  
    \begin{tabular}{lcc|cc}
	\toprule				
	{Number of}  & \tmc2{c}{HSJA} & \tmc2{|c}{Sign-Flip}    \\
        \cmidrule{2-5}
  	{Queries}    &   Recogn. & Detect. & Recogn. & Detect. \\
	\midrule
        16     &      91.7\% &  100.0\% &      85.1\% &  99.4\% \\
        32     &      90.8\% &  100.0\% &      87.4\% &  99.6\% \\	
        64     &      98.8\% &  100.0\% &      95.8\% &  99.5\% \\	
        128    &      99.9\% &  100.0\% &      98.3\% &  99.3\% \\	
        256    & \tb{100.0\%} &  100.0\% & \tb{99.8\%} &  99.8\% \\	
        \bottomrule                        
   \end{tabular}
  \caption{Effect of number of queries to form HoDS for GWAD prediction and detection performance over HSJA and Sign-Flip attacks. "Recogn." corresponds to the attack recognition, and "Detect." corresponds to the binary classification (benign/attack).}
  \label{Tab:effect_query_size}   
\end{table}
GWAD requires a set of queries to predict and detect query-based attack. In \Cref{Tab:effect_query_size}, we report the attack classification and detection performance of GWAD over various number of queries to make HoDS feature. GWAD monitors 10K attack queries from HSJA~\cite{HSJA} and Sign-Flip~\cite{SIGN_FLIP}. The classification success rate of GWAD is gradually improved as the number of queries to form HoDS grows reaching almost $100\%$ classification accuracy at 256 queries.

\end{document}